\documentclass[11pt]{article}
\usepackage{latexsym}  
\usepackage{amssymb}
\usepackage{graphicx}
\usepackage{amsmath}
\usepackage{mathrsfs}

\newcommand{\be}{\begin{equation}}
\newcommand{\ee}{\end{equation}}
\newcommand{\bea}{\begin{eqnarray}}
\newcommand{\eea}{\end{eqnarray}}
\newcommand{\bref}[1]{(\ref{#1})}
\newcommand{\bs}{\boldsymbol}

\newcommand{\bi}{\bibitem}
\newcommand{\pa}{\partial}

\newcommand{\bomega}{\boldsymbol{\Omega}}
\begin{document}
\begin{titlepage}
\today
\vspace{4\baselineskip}
\begin{center}
{\Large\bf Systematic Errors in General Spin Precession in Storage Ring}
\end{center}
\vspace{1cm}
\begin{center}
{\large Takeshi Fukuyama
\footnote{E-mail:fukuyama@se.ritsumei.ac.jp}}
\end{center}
\vspace{0.2cm}
\begin{center}
{\small \it Research Center for Nuclear Physics (RCNP),
Osaka University, Ibaraki, Osaka, 567-0047, Japan}\\[.2cm]



\vskip 10mm
\end{center}
\vskip 10mm

\begin{abstract}
We analyse the systematic errors on the spin precession of particles in the storage ring, especially in the muon g-2/EDM project. There particles have both anomalous magnetic moments (g-2) and elctric dipole moments(EDM), and their positions and velocity directions have both $O(10^{-3}-10^{-4})\equiv O(\epsilon)$ extensions. In order to measure these dipole moments up to $0.1$ppm or more we determine the precession frequency up to $O(\epsilon^2)$. Our analytical formulation includes the Farley's pitch correction in the special case and can be applied to more general experimental setups.
\end{abstract}
\end{titlepage}
\section{Introduction and the Review of Our Formulation}
In the recent papers, we discussed general spin precession and betatron oscillation in storage ring \cite{fuku1,fuku2}. "General" has the duplicate meanings. One is that the beam has a spread profile not only in vertical but also radial directions. Another is that it incorporates permanent electric dipole moment (EDM) as well as anomalous magnetic dipole moment (MDM) \cite{fuku3}. The measurement of both dipole moments are the smoking gun of the new physics beyond the standard model (SM) \cite{Ritz, fuku4}. 
Recently the frequency shifts induced by magnetic field gradients and ${\bf v}\times {\bf E}$ effects on the neutron EDM \cite{Pendlebury} and muon g-2 MDM have been discussed. In these situations, the analytical estimation of the systematic errors are very important. Unfortunately there is a controversy on the definition of the observed spin precession (Which is correct, \bref{Nelsonf} or \bref{Nelsonf2} ?). In this paper we show that \bref{Nelsonf2} is correct for the given experimental setup and reproduces the Farley's result \cite{Farley} in the special case. 

In order to explain the problem, let us start with the review of the previous work \cite{fuku2}. Imaging the J-PARC muon g-2/EDM experiment \cite{J-PARC}, we considered ${\bf E}=0$ in the latter part of the previous paper.
In this paper, we consider the case ${\bf E}\neq 0$ but with the constant magnitude of the velocity like the case of BNL-E821 \cite{Bennett} and its successor at FNAL. Also taking the strength of weak focusing ${\bf B}$ and ${\bf E}$ fields into consideration, we make an approximation at higher order than the previous paper with respect to the small extents. 

The kinetic energy (including the rest energy) satisfies
\be
\frac{dE_{kin}}{dt}=e{\bf E}\cdot {\bf v}
\label{kinetic}
\ee
with
\be
E_{kin}\equiv \frac{m}{\sqrt{1-v^2}}\equiv \gamma m
\ee
and constant magnitude of the velocity requires
\be
{\bf E}\cdot{\bf v}=0.
\label{Ev}
\ee
We do not assume this until it will be specified later. Here and hereafter we use the $\hbar=c=1$ units.
In the presence of both ${\bf B}$ and ${\bf E}$, the Lorentz equation is given by
\be
\gamma m \frac{d{\bf v}}{dt}\equiv \gamma m\dot{{\bf v}}= e\left({\bf E}+{\bf v}\times {\bf B}-{\bf v}({\bf v}\cdot{\bf E})\right).
\label{Lorentz1}
\ee
The Lorentz equation is also expressed as
\be
\frac{d(\gamma m {\bf v})}{dt}=e\left({\bf E}+{\bf v}\times {\bf B}\right).
\ee
The angular velocity of the spin rotation of the generalized Thomas-Bargman-Michel-Telegdi (BMT) equation in the laboratory system is given by \cite{fuku2}
\bea
\bomega_s&=&-\frac{e}{m}\left[\left(a+\frac{1}{\gamma}\right){\bf B}-\frac{\gamma a}{\gamma+1}({\bf v}\cdot{\bf B}){\bf v}-\left(a+\frac{1}{\gamma+1}\right){\bf v}\times{\bf E}\right.\nonumber\\
&+&\left.\frac{\eta}{2}\left({\bf E}-\frac{\gamma}{\gamma+1}({\bf v}\cdot{\bf E}){\bf v}+{\bf v}\times {\bf B}\right)\right].
\label{Nelsonh}
\eea
One usually considers the spin motion relative to the beam direction. It follows from \bref{kinetic} and \bref{Lorentz1} that the unit vector in direction of the velocity (momentum), ${\bf N}={\bf v}/v={\bf p}/p$ obeys
\be
\frac{d{\bf
N}}{dt}=\frac{\dot{\bf v}}{v}
-\frac{\bf v}{v^3}\left({\bf v}\cdot\dot{\bf v}\right)=\bomega_p\times{\bf N}, ~~~ \bomega_p=\frac{e}{m\gamma}\left(\frac{{\bf
N}\times{\bf E}}{v}-{\bf B}\right).
\ee 
In the particle rest frame (Frenet Serret coordinates), the unit vectors transform as
\bea
{\bf e}_1'&=& ~~~~~~~~\kappa{\bf e}_2\nonumber\\
{\bf e}_2'&=&-\kappa{\bf e}_1~~~~~~+\tau{\bf e}_3 \\
{\bf e}_3'&=&~~~~~~-\tau{\bf e}_2 \nonumber,
\label{FS}
\eea
where dash indicates derivative wrt $x$. 

Thus, the angular velocity of the spin rotation relative to the beam direction (particle rest frame) is given by
\be
\bomega'=\bomega_s-\bomega_p=-\frac{e}{m}\left[a{\bf B}-\frac{\gamma a}{\gamma+1}({\bf v}\cdot{\bf B}){\bf v}-\left(a-\frac{1}{\gamma^2-1}\right){\bf v}\times{\bf E}+\frac{\eta}{2}\left({\bf E}-\frac{\gamma}{\gamma+1}({\bf v}\cdot{\bf E}){\bf v}+{\bf v}\times {\bf B}\right)\right].
\label{Nelsonf}
\ee
However, it is the time variation of the deviation from the reference orbit that we measure by the detector around the planer reference orbit, which is described in the cylindrical coordinates along the refernce orbit (Figure 1).
They correspond to $\tau=0,~\kappa=-1/\rho$ in \bref{FS}.
The observed spin frequency in this frame is given by
\bea
\label{Nelsonf2}
&&\bomega=\bomega_s-\bomega_{pz}\nonumber\\
&&=-\frac{e}{m}\left[a{\bf B}-\frac{a\gamma}{\gamma+1}{\bf v}({\bf v}\cdot{\bf B})+\left(\frac{1}{\gamma^2-1}-a\right)({\bf v}\times\bs{E})\right.\\
&&\left. +\frac{1}{\gamma}\{\bs{B}_\parallel-\frac{1}{v^2}({\bf v}\times\bs{E})_\parallel\}+\frac{\eta}{2}(\bs{E}-\frac{\gamma}{\gamma+1}{\bf v}({\bf v}\cdot{\bf E})+{\bf v}\times\bs{B})\right].\nonumber
\eea
Here $\bomega_{pz}$ is the z-component of $\bomega_{p}$, and $\bs{B}_\parallel$ indicates the projected part of $\bs{B}$ onto the plane spanned by the reference orbit, that is,
\be
\bs{B}=(B_x,B_y,B_z)=(\bs{B}_\parallel, B_z),~~\bs{E}=(\bs{E}_\parallel, E_z).
\ee 

\begin{figure}[h]
\begin{center}
\includegraphics[scale=2.0]{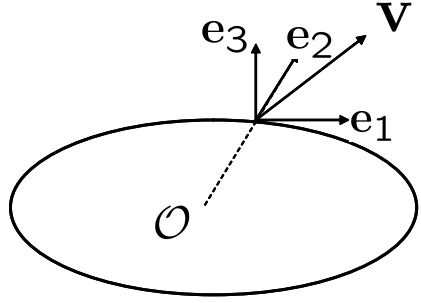}
\end{center}  
\caption{\label{fig:geom_config}
The configuration of betatron oscillation. ${\bf e}_1$(x-axis), ${\bf e}_2$(y-axis) coordinates are those projected on the averaged plane (horizontal plane) of the storage ring. ${\bf e}_3$(z-axis) is vertical to the horizontal plane. ${\bf v}$ is the particle velocity (emphasized to see the small pitch correction) which has both y, z components with main x component.}
\end{figure}

In this frame, 
\be
\frac{d{\bf e}_1}{dx}=-\frac{{\bf e}_2}{\rho},~~\frac{d{\bf e}_2}{dx}=\frac{{\bf e}_1}{\rho},~~\frac{d{\bf e}_3}{dx}=0.
\ee
Here ${\bf e}_1$ is the tangential unit vector (x-component) to the beam's averaged circular motion in the horizontal plane (spanned by x,y coordinates), 
${\bf e}_2$ is the radial unit vector (y-component) in the horizontal plane and ${\bf e}_3$ is the vertical unit vector (z-component) to the horizontal plane.
$\rho$ is the radius of the averaged circle and
\be
\frac{1}{\rho}=\left|\frac{eB_0}{\gamma mv}\right|.
\label{rho}
\ee 
Here we proceed to study a particle in the bunch whose position ${\bf r}$ is given by
\be
{\bf r}=(\rho+y){\bf e}_2+z{\bf e}_3
\label{profile}
\ee
with respect to the reference orbit. 

So let us consider how Eq.\bref{Nelsonf2} is expressed in these profiles \cite{Yokoya}. The velocity of the profile of \bref{profile} is
\be
{\bf v}=\frac{d{\bf r}}{dt}=\frac{dx}{dt}\frac{d{\bf r}}{dx}=\frac{dx}{dt}\left[\left(1+\frac{y}{\rho}\right){\bf e}_1+y'{\bf e}_2+z'{\bf e}_3\right],
\label{velocity}
\ee
where $'$ indicates the derivative with respect to $x$. Therefore, the absolute value $v$ of ${\bf v}$ is given by
\be
v=\frac{dx}{dt}\sqrt{\left(1+\frac{y}{\rho}\right)^2+y'^2+z'^2}\equiv \frac{dx}{dt}{\mathscr N}.
\ee
In the previous paper we proceeded the arguments under the condition of ${\bf E}=0$. In this paper we relax this condition as ${\bf E}$ is weak but $\neq 0$. 
One of the typical examples of focusing fields of ${\bf B}$ and ${\bf E}$ are
\be
{\bf B}=\left(0, ~gz,~ B_0+gy-\frac{g}{2\rho}z^2\right), ~~~~{\bf E}=(0,~ Ky, ~-Kz)
\label{focus}
\ee
with $g\equiv \frac{\pa B_z}{\pa y}$.
They satisfy 
\be
\nabla\times {\bf B}=0, ~~~~\nabla\times {\bf E}=0. 
\label{rotation}
\ee
We assume $y/\rho,~z/\rho,~y',~z'$ are small quantities of same order $\epsilon$. (In \bref{order} we will show that this is indeed the case.) and we take up to $O(\epsilon^2)$.
Hereafter we give the approximation formulae for the general ${\bf B}$ and ${\bf E}$, not assuming the special form of \bref{focus} until we will argue at \bref{constraint1} and thereafter.

Here we list the useful relations and approximations,
\bea
&&\frac{1}{\gamma v^2}=\frac{\gamma}{\gamma^2-1}, \nonumber\\
&&\frac{1}{{\mathscr N}}=1-\frac{y}{\rho}+\frac{y^2}{\rho^2}-\frac{{y'}^2+{z'}^2}{2}+O(\epsilon^3), \\
&&\frac{1}{{\mathscr N}^2}=1-\frac{2y}{\rho}+\frac{3y^2}{\rho^2}-{y'}^2-{z'}^2+O(\epsilon^3), \nonumber
\eea
and the divergence and rotation of a general vector ${\bf B}$ (and ${\bf E}$)
in this approximation are
\bea
\label{veq1}
&&{\bf v}\cdot {\bf B}=\frac{v}{{\mathscr N}}\left((1+\frac{y}{\rho})B_x+y'B_y+z'B_z\right)\nonumber\\
&&\approx v\left((1-\frac{y'^2+z'^2}{2})B_x+(1-\frac{y}{\rho})(y'B_y+z'B_z)\right)+O(\epsilon^3),\\
\label{veq2}
&& ({\bf v}\cdot {\bf B}){\bf v}\approx v^2\left[\left((1-(y'^2+z'^2))B_x+(1-\frac{y}{\rho})(y'B_y+z'B_z)\right){\bf e}_1+y'\left((1-\frac{y}{\rho})B_x+y'B_y+z'B_z\right){\bf e}_2\right. \nonumber\\
&&\left.+z'\left((1-\frac{y}{\rho})B_x+y'B_y+z'B_z\right){\bf e}_3\right],\\
&&{\bf v}\times {\bf B}=\frac{v}{{\mathscr N}}\left[\left(y'B_z-z'B_y\right){\bf e}_1+\left(z'B_x-(1+\frac{y}{\rho})B_z\right){\bf e}_2+\left((1+\frac{y}{\rho})B_y-y'B_x\right){\bf e}_3\right]\nonumber\\
\label{veq3}
&&\approx v\left[(1-\frac{y}{\rho})(y'B_z-z'B_y){\bf e}_1+\left((1-\frac{y}{\rho})z'B_x-(1-\frac{y'^2+z'^2}{2})B_z\right){\bf e}_2\right.\\
&&\left.+\left((1-\frac{y'^2+z'^2}{2})B_y-y'(1-\frac{y}{\rho})B_x\right){\bf e}_3\right]. \nonumber
\eea
Hereafter we assume \bref{Ev} explicitly.
Substituting \bref{veq1}-\bref{veq3} into \bref{Nelsonf2}, we obtain
\bea
\label{Nelsonf4}
&&\bomega=-\frac{e}{m}\left[\left\{\frac{1}{\gamma}(a+1)B_x{\bf e}_1+\left((a+\frac{1}{\gamma})B_y-\frac{\eta}{2}vB_z\right){\bf e}_2+\left(aB_z+\frac{\eta}{2}vB_y\right){\bf e}_3\right\}\right. \nonumber\\ 
&&+\left\{\left(a(1-\frac{1}{\gamma})((y'^2+z'^2)B_x-(1-\frac{y}{\rho})(y'B_y+z'B_z))-(a+\frac{1}{\gamma+1})v(1-\frac{y}{\rho})(y'E_z-z'E_y)\right. \right.\nonumber\\
&&\left.\left.+\frac{\eta}{2}(E_x+v(1-\frac{y}{\rho})(y'B_z-z'B_y))\right){\bf e}_1\right.\\
&&\left.\left.+\left(-a(1-\frac{1}{\gamma})((1-\frac{y}{\rho})B_x+y'B_y+z'B_z)y'-(a+\frac{1}{\gamma+1})v((1-\frac{y}{\rho})z'E_x-(1-\frac{y'^2+z'^2}{2})E_z)\right.\right.\right.\nonumber\\
&&\left.\left.\left.+\frac{\eta}{2}(E_y+v((1-\frac{y}{\rho})z'B_x+\frac{y'^2+z'^2}{2}B_z))\right){\bf e}_2\right.\right.\nonumber\\
&&\left.\left.+\left(-a(1-\frac{1}{\gamma})((1-\frac{y}{\rho})B_x+y'B_y+z'B_z)z'+(\frac{1}{\gamma^2-1}-a)v((1-\frac{y'^2+z'^2}{2})E_y-y'(1-\frac{y}{\rho})E_x)\right.\right.\right.\nonumber\\
&&\left.\left.\left.+\frac{\eta}{2}(E_z-v(\frac{y'^2+z'^2}{2}B_y+y'(1-\frac{y}{\rho})B_x))\right){\bf e}_3\right\}\right].\nonumber
\eea
Similarly substituting \bref{veq1}-\bref{veq3} into \bref{Nelsonf}, we obtain
\bea
\label{Nelsonf5}
&&\bomega'=-\frac{e}{m}\left[\left\{\frac{a}{\gamma}B_x{\bf e}_1+\left(aB_y-\frac{\eta}{2}vB_z\right){\bf e}_2+\left(aB_z+\frac{\eta}{2}vB_y\right){\bf e}_3\right\}\right. \nonumber\\ 
&&+\left\{\left(a(1-\frac{1}{\gamma})((y'^2+z'^2)B_x-(1-\frac{y}{\rho})(y'B_y+z'B_z))-(a-\frac{1}{\gamma^2-1})v(1-\frac{y}{\rho})(y'E_z-z'E_y)\right. \right.\nonumber\\
&&\left.\left.+\frac{\eta}{2}(E_x+v(1-\frac{y}{\rho})(y'B_z-z'B_y))\right){\bf e}_1\right.\\
&&\left.\left.+\left(-a(1-\frac{1}{\gamma})((1-\frac{y}{\rho})B_x+y'B_y+z'B_z)y'-(a-\frac{1}{\gamma^2-1})v((1-\frac{y}{\rho})z'E_x-(1-\frac{y'^2+z'^2}{2})E_z)\right.\right.\right.\nonumber\\
&&\left.\left.\left.+\frac{\eta}{2}(E_y+v((1-\frac{y}{\rho})z'B_x+\frac{y'^2+z'^2}{2}B_z))\right){\bf e}_2\right.\right.\nonumber\\
&&\left.\left.+\left(-a(1-\frac{1}{\gamma})((1-\frac{y}{\rho})B_x+y'B_y+z'B_z)z'-(a-\frac{1}{\gamma^2-1})v((1-\frac{y'^2+z'^2}{2})E_y-y'(1-\frac{y}{\rho})E_x)\right.\right.\right.\nonumber\\
&&\left.\left.\left.+\frac{\eta}{2}(E_z-v(\frac{y'^2+z'^2}{2}B_y+y'(1-\frac{y}{\rho})B_x))\right){\bf e}_3\right\}\right].\nonumber
\eea
In \bref{Nelsonf4} and \bref{Nelsonf5}, the first lines are the dominant contributions and the others are corrections. Specially important terms are $aB_z{\bf e}_3$ and $-\frac{\eta}{2}vB_z{\bf e}_2$ in the dominant contributions and $-a(1-\frac{1}{\gamma})z'^2B_z{\bf e}_3$ in the corrected terms.

If we do not assume \bref{Ev}, we may simply add the following term in $\eta/2$ term of \bref{Nelsonf4} and \bref{Nelsonf5},
\bea
&&-\frac{\gamma}{\gamma+1}({\bf v}\cdot {\bf E}){\bf v}\nonumber\\
&&\approx -\frac{\gamma}{\gamma+1}v^2\left[\left((1-(y'^2+z'^2))E_x+(1-\frac{y}{\rho})(y'E_y+z'E_z)\right){\bf e}_1+y'\left((1-\frac{y}{\rho})E_x+y'E_y+z'E_z\right){\bf e}_2\right.\nonumber\\
&&\left.+z'\left((1-\frac{y}{\rho})E_x+y'E_y+z'E_z\right){\bf e}_3\right].
\eea
The difference of \bref{Nelsonf4} and \bref{Nelsonf5} will be discussed in detail in the case of vertical oscillation in the next section.

Next, we must study the equation of motions of $y$ and $z$ to see the time variation of \bref{Nelsonf4}.
Constant magnitude of velocity indicates that
\be
\frac{dv}{dt}=\frac{d^2x}{dt^2}{\mathscr N}+{\mathscr N}'\left(\frac{dx}{dt}\right)^2=0
\ee
and, therefore,
\bea
\label{acceleration1}
&&\frac{d{\bf v}}{dt}=v^2\left[\left\{(1-\frac{2y}{\rho})\frac{y'}{\rho}-y'y''-z'z''\right\}{\bf e}_1+\left\{(1-\frac{2y}{\rho})y''-\frac{1}{\rho}(1-\frac{y}{\rho}+\frac{y^2}{\rho^2}-z'^2)\right\}{\bf e}_2 \right.\nonumber\\
&&\left. -\left\{\frac{y'z'}{\rho}-z''(1-\frac{2y}{\rho})\right\}{\bf e}_3\right].
\eea
Hereafter we consider the case of \bref{focus}.
This form together with \bref{Ev} and \bref{velocity} gives
\be
y^2-z^2=const.
\label{constraint1}
\ee
Consequently, in the $O(\epsilon^2)$ approximation, it follows from \bref{focus}, \bref{acceleration1} and from the Lorentz equation \bref{Lorentz1} with $\dot{\gamma}=0$ that
\bea\label{eq1}
&& y'\left(y''+\frac{1-n}{\rho^2}y\right)+z'\left(z''+\frac{n}{\rho^2}z\right)=0,\\
&&y''+\frac{1-n'}{\rho^2}y=\frac{1}{\rho}\left(2yy''+\frac{y^2}{\rho^2}+\frac{y'^2-z'^2}{2}\right), \label{eq2}\\
&&z''+\frac{n'}{\rho^2}z=\frac{1}{\rho}(2yz''+y'z')
\label{eq3}.
\eea
Here
\be
n\equiv -\frac{\rho}{B_0}\frac{\partial B_z}{\partial y},~~n'\equiv n\left(1-\frac{K}{v\partial B_z/\partial y}\right).
\ee
In \bref{eq1} we can replace $n$ by $n'$ by virtue of \bref{constraint1}.
Thus the well knowm Hill's equations,
\bea
&&y''+\frac{1-n'}{\rho^2}y=0, \label{eqmotion1}\\
&&z''+\frac{n'}{\rho^2}z=0,
\label{eqmotion2}
\eea
is valid in $O(\epsilon)$ at first glance. Fortunately the right-hand parts of \bref{eq2} and \bref{eq3} do not act in \bref{Nelsonf4} and \bref{Nelsonf5} in the $O(\epsilon^2)$ because these terms appear at most $O(\epsilon^3)$ in \bref{Nelsonf4} and \bref{Nelsonf5}. This is the same for \bref{constraint1}. Thus our formulation with \bref{eqmotion1} and \bref{eqmotion2} and without \bref{constraint1} are consistent up to $O(\epsilon^2)$. 
In the presence of momentum dispersion, \bref{eqmotion1} is modified as
\be
y''+\frac{1-n'}{\rho^2}y=\frac{1}{\rho}\frac{\delta p}{p},
\label{eqmotion3}
\ee
whereas \bref{eqmotion2} is not altered even in this case. The solutions of \bref{eqmotion3} and \bref{eqmotion2} are as follow \cite{Conte}.
\be
\left(
  \begin{array}{c}
y\\
y'\\
\frac{\delta p}{p}\\
\end{array}
\right)=\left(
\begin{array}{ccc}
\cos \sqrt{1-n}\theta & \frac{\rho}{\sqrt{1-n}}\sin\sqrt{1-n}\theta &\frac{\rho}{1-n}(1-\cos\sqrt{1-n}\theta )\\
   -\frac{\sqrt{1-n}}{\rho}\sin\sqrt{1-n}\theta & \cos\sqrt{1-n}\theta & \frac{1}{\sqrt{1-n}}\sin \sqrt{1-n}\theta\\
0& 0& 1 \\
\end{array}
\right)
\left(
\begin{array}{c}
y_0\\
y_0'\\
\left(\frac{\delta p}{p}\right)_0\\
\end{array}
\right)
\label{sol1}
\ee
Here $y_0,~y_0', \left(\frac{\delta p}{p}\right)_0$ are $y(0),~y'(0), \frac{\delta p(0)}{p(0)}$, respectively, and $\theta=x/\rho$. Similarly,
\be
\left(
  \begin{array}{c}
z\\
z'\\
\end{array}
\right)=
\left(
\begin{array}{cc}
\cos \sqrt{n}\theta & \frac{\rho}{\sqrt{n}}\sin\sqrt{n}\theta \\
   -\frac{\sqrt{n}}{\rho}\sin\sqrt{n}\theta & \cos\sqrt{n}\theta\\
\end{array}
\right)
\left(
\begin{array}{c}
z_0\\
z_0'\\ 
\end{array}
\right).
\ee

In conclusion of this section, we have calculated the spin precession to $O(\epsilon^2)$. This is essential because Farley's pitch correction appears in this order \cite{Farley}. Its main contribution cooresponds to $-a(1-\frac{1}{\gamma})z'^2B_z$ of ${\bf e}_3$ component in \bref{Nelsonf4}, supplemented by those of $x,y$ components. These are explained in the next section.
Our formula is the generalization of his formula extended to $y,y',z,z'$ and to the presence of the electric dipole moment. This is crucial especially in the presence of electric dipole moment since its main contribution appers in ${\bf e}_2$ direction.

\section{The Application of Our Formulation to the Farley's Setup}
The Farley's pitch correction is well known and accepted among wide communities. Whereas our formulation, especially on the definition of $\Omega$ is not so well understood yet. In this section we show that our formulation reproduces fully the Farley's results analytically in his setup. 

It follows from \bref{Nelsonf2} that
\be
\Omega_z=-\frac{e}{m}\left[aB_z-\frac{a\gamma}{\gamma +1}v_z(vB_v)\right],
\ee
where $B_v$ is the projection of $\bs{B}$ on moving particle's direction and therefore\footnote{$z'=\frac{dz}{dx}=\tan\psi\approx \psi$. It follows from \bref{eqmotion2} that $\omega_p=\sqrt{n}\frac{eB_0}{\gamma m}$ since $\theta=\Omega_p t=\frac{eB_0}{\gamma m}t$ in the absence of ${\bf E}$.}
\be
B_v=B_z\sin\psi,~~v_z=v\sin \psi. 
\ee
Then
\be
\Omega_z=-\frac{e}{m}aB_z\left(1-\frac{\gamma}{\gamma+1}v^2\sin^2\psi\right)=\Omega_0\left(1-\frac{\gamma-1}{\gamma}\psi^2\right).
\label{pz}
\ee
Here $\Omega_0\equiv \frac{|e|}{m}aB_z$ and
\be
\sin \psi\approx \psi=\psi_0\sin (\omega_pt).
\ee
Aso
\be
\Omega_x=-\frac{e}{m}\left( -\frac{a\gamma}{\gamma+1}v^2B_v\right)=-\Omega_0\frac{\gamma-1}{\gamma}\sin \psi.
\label{px}
\ee
We are considering weak magnetic focusing \bref{focus}, $B_x=0,~B_y\neq 0$ and
\be
\dot{p}_z=e(\bs{v}\times \bs{B})_z=evB_y.
\ee
Since
\be
p_z=\gamma mv\sin\psi\equiv p\sin\psi
\ee
and
\be
\dot{p}_z=p\dot{\psi}\cos\psi\approx p\dot{\psi}=p\omega_p\psi_0\cos(\omega_pt)=evB_p.
\ee
Then we obtain
\be
eB_y=\gamma m\omega_p\psi_0\cos(\omega_pt).
\ee
So it goes from \bref{Nelsonf2} with the presence of weak focusing \bref{focus} that
\bea
&&\Omega_y=-\frac{e}{m}\left(aB_y+\frac{1}{\gamma}B_y\right)\equiv -\frac{e}{m\gamma}fB_y\nonumber\\
&&=-f\omega_p\psi_0\cos(\omega_pt)
\label{py}
\eea
with
\be
f\equiv (1+a\gamma).
\label{focus1}
\ee
It should be remarked that \bref{Nelsonf} gives $f=a\gamma$. \footnote{Likewise, for electric focusing we obtain
\be
\Omega_y=-\frac{eE_z}{m\gamma v}\left(1+\gamma v^2a-\frac{1}{\gamma}\right)
\ee
and 
\be
f=1+\gamma v^2a-\frac{1}{\gamma}.
\label{focus2}
\ee
Eqs. \bref{focus1} and \bref{focus2} show that our ${\bf \Omega}$ (and not ${\bf \Omega}'$) coincides with that of Farley \cite{Farley}.}

Thus we obtain
\bea
&&\bs{\Omega}=\Omega_0\left[\left(1-\frac{\gamma-1}{\gamma}\psi_0^2\sin^2(\omega_pt)\right)\bs{e}_3\right.\nonumber\\
&&\left.-\frac{\gamma-1}{\gamma}\psi_0\sin(\omega_pt)\bs{e}_1-f\omega_p\psi_0\cos(\omega_pt)\bs{e}_2\right].
\label{p2}
\eea
So far we have assumed that $y/\rho,~z/\rho$ are of same orders as $y',~z'$.
Let us consider whether it is indded the case in the case of weak magnetic focusing. The relative deviation of ${\bf \Omega}$ is divided into two parts,
\be
\langle\frac{\delta \Omega}{\Omega}\rangle =\langle\frac{\delta \Omega}{\Omega}\rangle _{\delta B=0}+\langle\frac{\delta \Omega}{\Omega}\rangle _{z'=0}.
\ee
Here and hereafter $\langle ...\rangle$ indicates the time averaging. 
Since, from \bref{sol1}, 
\be
y=y_0\cos\sqrt{1-n}\theta+y_0'\frac{\rho}{\sqrt{1-n}}\sin\sqrt{1-n}\theta+\frac{\rho}{1-n}(1-\cos\sqrt{1-n}\theta)\left(\frac{\delta p}{p}\right)_0,
\ee
we obtain 
\be
\langle y\rangle=\frac{\rho}{1-n}\left(\frac{\delta p}{p}\right)_0=\frac{\rho}{1-n}(\cos\psi_0-1)=-\frac{\rho}{1-n}\frac{\psi_0^2}{2},
\label{farley2}
\ee
\be
\langle z^2\rangle =\langle (z_0\cos \sqrt{n}\theta+z_0'\frac{\rho}{\sqrt{n}}\sin\sqrt{n}\theta)^2\rangle=\frac{\rho^2}{n}\frac{z_0'^2}{2}=\frac{\rho^2}{n}\frac{\psi_0^2}{2}.
\label{farley3}
\ee
In the second equality we have used $z_0=0$ due to the Farley's situation \cite{Farley}, and also we obtain
\be
\left|\frac{z}{\rho}\right|=\left|\frac{z'}{\sqrt{n}}\right|.
\label{order}
\ee
It follows from Eqs. \bref{focus}, \bref{Nelsonf4}, and \bref{Farley} that
\bea
\label{farley4}
&&\langle\frac{\delta \Omega}{\Omega}\rangle _{z'=0}=\frac{1}{B_0}g\langle y\rangle-\frac{1}{B_0}g\frac{\langle z^2\rangle}{2\rho}=\frac{n}{1-n}\frac{\psi_0^2}{2}+\frac{\psi_0^2}{4}, \\
&&\langle\frac{\delta \Omega}{\Omega}\rangle _{\delta B=0}=-\langle \frac{z'^2}{2}\rangle\rangle=-\frac{\psi_0^2}{4}.
\label{farley5}
\eea
Here $z'^2$ term is reduced by factor two by the contribution of $\Omega_x,~\Omega_y$ as will be shown in \bref{Farley}.
Finally we obtain \cite{Yannis}
\be
\langle\frac{\delta \Omega}{\Omega}\rangle =\frac{n}{1-n}\frac{\psi_0^2}{2}.
\ee
We need still more step to get the observed spin precession. The following arguments are due to Farley \cite{Farley} and we will write down explicitly since the original paper had typos.

Combining Eqs. \bref{pz},\bref{px}, and \bref{py}. we obtain the spin motion as follows:
\be
\frac{\bs{s}}{s}\equiv \bs{\xi}=(\cos\theta\cos\phi,\cos\theta\sin\phi,\sin\theta).
\ee
Then
\be
\frac{d\bs{\xi}}{dt}=\bs{\Omega}\times \bs{\xi}.
\label{xi}
\ee
The z component of \bref{xi} gives
\be
\dot{\theta}=\Omega_x\sin\phi-\Omega_y\cos \phi=-\Omega_0\frac{\gamma-1}{\gamma}\sin\psi\sin\phi+f\omega_p\psi_0\cos(\omega_pt)\cos\phi.
\label{xi2}
\ee
The x componenent of \bref{xi} is
\be
\frac{d\xi_x}{dt}=-\dot{\theta}\sin\theta\cos\phi-\dot{\phi}\cos\theta\sin\phi=\Omega_y\sin\theta-\Omega_z\cos\theta\sin\phi
\label{xi3}.
\ee
Substituting \bref{xi2} for $\dot{\theta}$, we obtain
\bea
&&-\dot{\phi}\cos\theta\sin\phi=\sin\theta\cos\phi(\Omega_x\sin\phi-\Omega_y\cos\phi)+\Omega_y\sin\theta-\Omega_z\cos\theta\sin\phi\nonumber\\
&&=-\Omega_z\cos\theta\sin\phi+\Omega_x\sin\theta\sin\phi\cos\phi-\Omega_y\sin\theta(\cos^2\phi-1).
\label{xi4}
\eea
Hence
\bea
\label{xi5}
&&\dot{\phi}=\Omega_z-\Omega_x\tan\theta\cos\phi-\Omega_y\tan\theta\sin\phi\\
&&=\Omega_0\{1-\frac{\gamma-1}{\gamma}\psi_0^2\sin^2(\omega_pt)\}+\Omega_0\psi_0\frac{\gamma-1}{\gamma}\tan\theta\cos\phi\sin(\omega_pt)+\omega_p\psi_0f\tan\theta\sin\phi\cos(\omega_pt).\nonumber
\eea
The solution of \bref{xi4} is
\be
\theta=\frac{A_+}{\Omega_0+\omega_p}\sin\{(\Omega_0+\omega_p)t+\xi\}-\frac{A_-}{\Omega_0-\omega_p}\sin\{(\Omega_0-\omega_p)t+\xi\},
\label{thetasol}
\ee
where
\be
A_\pm\equiv \frac{1}{2}\psi_0\{\Omega_0\frac{\gamma-1}{\gamma}\pm f\omega_p\}.
\ee
Thus we can rewrite Eq.\bref{p2}
\be
\bs{\Omega}=\left(a_0+a_3\cos(2\omega_pt)\right)\bs{e}_3+a_2\cos(\omega_pt)\bs{e}_2+a_1\sin(\omega_pt)\bs{e}_1,
\label{p3}
\ee
where $a_i$ are constants given by
\bea
a_0&\equiv& \Omega_0\left(1-\frac{\gamma-1}{2\gamma}\psi_0^2\right),~~
a_1\equiv -\Omega_0\frac{\gamma-1}{\gamma}\psi_0,\\
a_2&\equiv&-f\omega_p\psi_0,~~a_3\equiv \Omega_0\frac{\gamma-1}{2\gamma}\psi_0^2\nonumber
\label{p4}
\eea
Here 
\be
a_0=O(1),~~a_1,~a_2=O(\epsilon),~~a_3=O(\epsilon^2)
\ee
and we can solve \bref{p2} in terms of spin components.
Substituting \bref{thetasol} into \bref{xi5}, we obtain
\be
\dot{\phi}=\Omega_0\left(1-\frac{\gamma -1}{2\gamma}\psi_0^2\right)+\frac{A_+^2}{2(\Omega_0+\omega_p)}+\frac{A_-^2}{2(\Omega_0-\omega_p)}+\mbox{oscillatong terms.}
\label{phidot}
\ee
Here we have used
\be
\langle \sin\Omega_0t\sin\phi\rangle=\langle \cos\Omega_0t\cos\phi\rangle=\frac{1}{2}
\ee
since $\phi\approx \Omega_0 t$.
The observed time averaged spin precession is given by the time averaging of $\dot{\phi}$, which is different from either $\Omega_0$ or $\Omega_z$ (first term of \bref{phidot}), and
\be
\langle \dot{\phi}\rangle=\Omega_o(1-C)
\label{Farley}
\ee
where
\be
C=\frac{1}{4}\psi_0^2\left[1-\frac{\Omega_0^2}{\gamma^2(\Omega_0^2-\omega_p^2)}-\frac{\omega_p^2(f-1)(f-1+\frac{2}{\gamma})}{\Omega_0^2-\omega_p^2}\right].
\label{farley}
\ee
Thus, apart from the resonance, the pitch correction is reduced by factor 2, namely from $-\frac{1}{2}\psi_0^2\to-\frac{1}{4}\psi_0^2$, by means of the contributions of $\Omega_x,~\Omega_y$. There $f\approx 1$ is essential (See \bref{focus1}).
\section{Discussion}
We have developed the analytical estimation of systematic errors in muon spin precession up to $O(\epsilon^2)$.
It has been shown that our formulation reproduces the well known Farley's pitch correction in the special case.
Moreover, it includes more general case than that of Farley.
Indee, the injected beam has some extended profiles in both radial and vertical directions.
In deriving Eq.\bref{farley3} we have set $z_0$ equals to zero. However in the realistic case neither $y_0$ nor $z_0$ is zero, and 
\be
\langle z^2\rangle =\frac{z_0^2}{2}+\frac{\rho^2}{n}\frac{z_0'^2}{2}.
\ee
The second term of \bref{farley4} does not cancel out with \bref{farley5} in such general cases. Thus we are required to match our formulation with more complicated situation in advancing to higher precision level.
Analytical studies developed by us will play very important roles in these situations together with numerical error estimation programming.

\section*{Acknowledgments}

This work is supported in part by 
  Grant-in-Aid for Science Research from the Ministry of Education, Science and Culture No.~17H01133.

\end{document}